# Sensing as a Service and Big Data

Arkady Zaslavsky[#1], Charith Perera[#*2], Dimitrios Georgakopoulos[#3]

[#]ICT Centre, CSIRO, Canberra, ACT, 2601, Australia

[1]arkady.zaslavsky@csiro.au
[2]charith.perera@csiro.au
[3]dimitrios.georgakopoulos@csiro.au

[*]Research School of Computer Science, The Australian National University,
Canberra, ACT 0200, Australia

*Abstract*— **Internet of Things (IoT) will comprise billions of devices that can sense, communicate, compute and potentially actuate. Data streams coming from these devices will challenge the traditional approaches to data management and contribute to the emerging paradigm of big data. This paper discusses emerging Internet of Things (IoT) architecture, large scale sensor network applications, federating sensor networks, sensor data and related context capturing techniques, challenges in cloud-based management, storing, archiving and processing of sensor data.**

## 1. Introduction

The modern technology-savvy world is full of devices comprising sensors, actuators, and data processors. Such concentration of computational resources enables sensing, capturing, collection and processing of real time data from billions of connected devices serving many different applications including environmental monitoring, industrial applications, business and human-centric pervasive applications. These developments have brought us to the era of Internet of Things (IoT) [1] by introducing IoT in 1998as concept [2]. However, sensing the environment around us and objects populating this environment became synonymous with the introduction of pervasive or ubiquitous computing by the paper 'The Computer for 21st Century' [3] in 1991 in the same year where World Wide Web became available. The major enabler of IoT is sensor networks. IoT has three unique features [4]: intermittent sensing, regular data collection, and Sense-Compute-Actuate (SCA) loops.

In 2010, the total amount of data on earth exceeded one zettabyte (ZB) [5]**,** [6] (see figure 1). By end of 2011, the number grew up to 1.8 ZB [7]. Further, it is expected that this number will reach 35 ZB in 2020. As in many cases with ICT, this estimate may prove to be too conservative.

IEEE Spectrum [6] recognises both sensors and big data as to of the five technologies that will shape the world (figure 2). According to Gartner Research [8], by 2015, wirelessly networked sensors in everything we own will form a new Web. But it will only be of value if the 'terabyte' of data it generates can be collected, analysed and interpreted. Further, European Commission [7] predicts that the present 'Internet of PCs' will move towards an 'Internet of Things' in which 50 to 100 billion devices will be connected to the Internet by 2020.

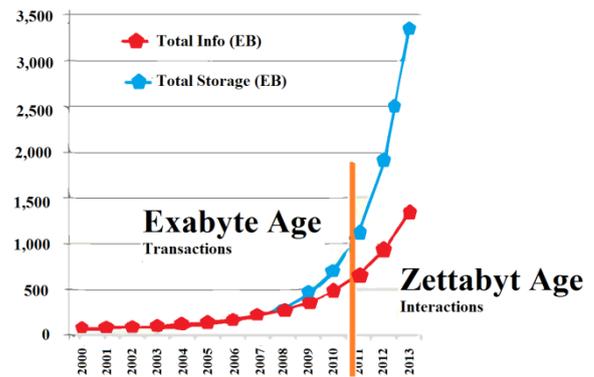

Figure 1: The total amount of data generated on earth exceeded one zettabyte in 2010.It is predicted that data volume will grow exponentially as depicted[1].

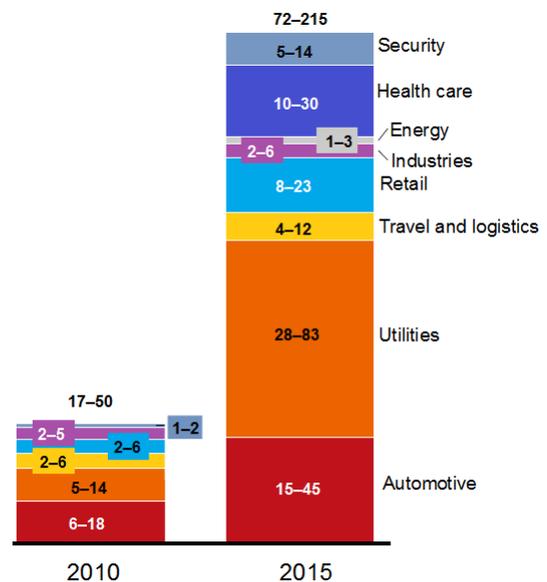

Figure 2: Data generated from the Internet of Things will grow exponentially as the number of connected nodes increases. Estimated numbers of connected nodes based on different sectors are presented in Millions [9].

---

[1] www.teradata.com

[7] defines IoT as the growing and largely invisible web of interconnected smart objects that promises to transform the way we interact with everyday things.

Due to recent development in sensor devices and other related technologies, the cost of data acquisition has dramatically come down [10], [11]. Even though sometime, initial costs are high, continues data acquisition remains very cheap. Initial costs are also going down with recent developments in sensor technologies such as Arduino (arduino.com).

According to the BCC Research [12], global market for sensors was around $56.3 billion in 2010. In 2011, it was around $62.8 billion. Global market for sensors is expected to increase up to $91.5 billion by 2016, at a compound annual growth rate of 7.8%.

IBM attributed the growing amount of big data towards instrumented, interconnected, intelligent world which is envisioned by Internet of Things [13]. Infosys [14] acknowledges intelligence, cloud computing, and sensor networks as three significant themes in the evolution of pervasive computing.

2. **Sensing Big Data**

Big data is not new concept or idea. However, earlier notions of big data was limited to few organizations such as Google, Yahoo, Microsoft, and European Organization for Nuclear Research (CERN). However, with recent developments in technologies such as sensors, computer hardware and the Cloud, the storage and processing power increase and the cost comes down rapidly. As a result, many sources (sensors, humans, applications) start generating data and organizations tend to store them for long time due to inexpensive storage and processing capabilities. Once that big data is stored, a number of challenges arise such as processing and analysing. Thus big data has become a buzz word in industry.

There is no clear definition for 'Big Data'. It is defined based on some of its characteristics. The big data does not mean the size. There are three characteristics that can be used to define big data, as also known as 3V's [13]: volume, variety, and velocity (figure 3).

- **Volume**: Volume relates to size of the data such as terabytes (TB), petabytes (PB), zettabytes (ZB), etc.
- **Variety**: Variety means the types of data. In addition, difference sources will produce big data such as sensors, devices, social networks, the web, mobile phones, etc. For example, data could be web logs, RFID sensor readings, unstructured social networking data, streamed video and audio.
- **Velocity**: This means how frequently the data is generated. For example, every millisecond, second, minute, hour, day, week, month, year. Processing frequency may also differ from the user requirements. Some data need to be processed real-time and some may only be processed when needed. Typically, we can identify three main categories: occasional, frequent, and real-time.

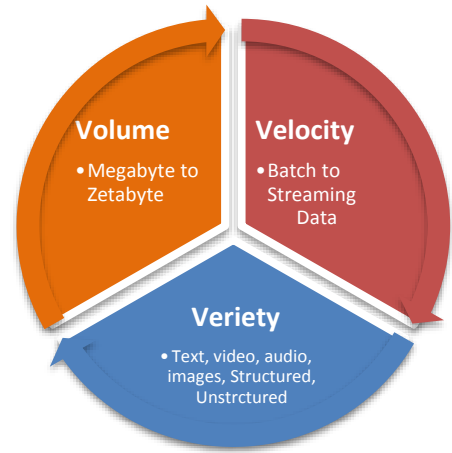

Fig 3: Characteristics of Big Data ($V^3$) [13]

Some researchers consider Value also as a main characteristic of big data. This means that somewhere within that data, there is some valuable information – golden data to extract, though most of the pieces of data individually may seem valueless.

EMC [15] has defined *big data as any attribute that challenges constraints of a system capability or business need.* For example, EMC recognizes 40MB PowerPoint presentation as a big data, because 40MB is big compared to typical size of a PowerPoint presentation. Further, 1PB animation and a 1TB medical image are considered as big data as they are big in size compared to the typical size of each. Further, the data we consider big today may not be considered big tomorrow due to the advances in processing, storage and other system capabilities.

There is a number of challenges related to big data. Due to its characteristics some of the inherited challenges in big data are capture, storage, search, analysis, and virtualization.

Some of the typical areas that produce big data are meteorology, genomics, physics, simulations, biology and environmental science. It is predicted that vehicles will be a major source that would produce big data [16]. The number of vehicles used and the growing rates around the world justify the prediction. Further, some of the potential application areas of big data analytics are smart homes, finance, log analysis, security, traffic control, telecommunications, search quality, manufacturing, trade analysis, fraud and risk [5], [17].

Big data is everywhere, even in jogging. One example of application of sensing technologies in our everyday life is *Nike+iPod/iPhone* application [18]. It is an application that collects and tracks information such as workout details, distance, calories burnt, etc. of a jogger using Nike footwear and iPhone/iPod. Another similar application is *iSmoothRun* (www.ismoothrun.com ). Further, it allows uploading data to the fitness social networks such as RunKeeper.com. The data becomes 'big' when we consider millions of users.

The state of California [19] has developed a greenhouse gas sensor network located throughout California where it collects massive amounts of real-time information about greenhouse gases and their behaviour.

The St. Anthony Falls Bridge in Minneapolis [13] consists of over 200 sensors embedded in strategic points. These sensors provide fully detailed sensor readings regarding the bridge to a monitoring system. This system collects variety of different measurements such as a change in temperature and the bridge's concrete reaction to that change is available for analysis.

Sensors collect information in natural disaster situations in order to optimally utilize resources and manage supply chains [5]. For example, during the disaster at Japan's Fukushima Daichi nuclear plant in 2011, sensors were used to take radiation measurements without human intervention [6].

The following statistics can be used to further explain the big data. For example, more than 12TB (155 million tweets) of tweets in Twitter and 25TB of log data (30 billion pieces of content) in Facebook are generated every day. Further, 30 billion RFID tags and 4.6 billion camera phones are used around the world today. In addition, 200 million smart meters to be operated in 2014. Moreover, there were 2 billion people on web in 2011 [5]**,** [13].

Four detectors in Large Hadron Collider (LHS) produced 13 petabyte (PB) of data in 2010 [20]**.** Wal-Mart generates 2.5 petabyte (PB) of customer transaction data every hour**.** Further, if Wal-Mart operates RFID on item level, it is expected to generate 7 terabytes (TB) of data every day [21].

In future, large radio astronomy arrays, for example, Square Kilometre Array (SKA), will generate data at rates far higher than that can be processed or stored affordably using current technology [17].

Boeing Jet [22] generates 10 terabytes (TB) of data per engine every 30 minutes. A single six hour flight would generate 240 terabyte (TB) of data. In addition, there are about 28537 commercial flights in the sky in United States on any given day. These statistics allow us to understand the scale of the sensor data generated. However, these collected data may not be processed all the time. Usually the information are closely analysed only when airplane get crashed. The big data has become an industry itself which is worth around USD 100 billion and growing at a 10% rate annually [23], where major players such as Oracle, Microsoft, SAP, and IBM spend significantly due to the potential value of big data.

Ninja Blocks (ninjablocks.com) can be identified as latest development in where IoT and cloud computing is merging together. Ninja Blocks are small and cheap devices which encompass variety of different sensors such as acceleration, temperature, current, humidity, motion, distance, sound, light and even capture video. In addition, these block are capable of connect themselves to the Ninja cloud where the data can be processes and stored. It has the capability to identify some of the event occur based on the sensor data and perform actions such as tweet, sms, email, Facebook or Dropbox uploading . Another similar start-up is i-voltmeter (i-voltmeter.com). These types of devices have the potential to generate big data in coming decade.

SFpark (sfpark.org) has deployed sensors over 8200 parking space in several San Francisco neighbourhoods in order to track real-time inventory and to perform changes to the pricing based on the consumer behaviour. School in Vittoria da Conquista, Brazil [24] has equipped uniforms with RFID chips allowing their parent to verify that their children are show up at school. Very large scale sensor networks domain [25] use cloud computing to process data in the cloud. Such data can be characterised as polymorphous, heterogeneous, large in quantity and time-limited. In a very large scale sensor network, manage the sensing resources and computational resources, and store and process these data are identified as key challenges. The potential applications are vehicle tracking, traffic dispatching, and expressway planning.

London [26] has deployed significant amount of sensor roadways in order to monitor traffic in real-time to optimum traffic management targeting 2012 Olympics. In general, sensors in roadways can be used to optimize route for fuel efficiency. Let's assume that governments deploy and own these sensors. However, there would be several other organizations that would interest on the data collected by those roadway sensors such as traffic update providers and weather information providers [22]. Government may deploy roadways sensors with the focus of collecting traffic information. However, several other organizations would like to access those sensor data so they can analyse sensor data in different perspectives and provide their customer with value added services. Further, these third party organizations can pay a fee to the government and get access to the sensor data provided by roadway sensors. In summary, government can provide sensing data as a service. Therefore, this Sensing as a service model can be applied well here.

The map in figure 4 shows how the big data have been generated and stored across the world.

3. **Why Big Data?**

Big data is important for us in many perspectives. The significant amount of data generated allows us to make decisions in timely manner where money can be saved and operations can be more optimized in both public and private sector. For example, in retail business, consumer behaviour and preferences can be understand by analysing the big data which includes, customer movement in the store or online webs site, transactions, product searches, etc. [27]. Big data allows *Data-Driven Decision Making* [28]. United States Healthcare Big Data World [29], which comprises records of over 50 million patients, uses data-driven concept to find out the challenges in healthcare sector. The potential queries that they are expecting to process over big data are very complex.

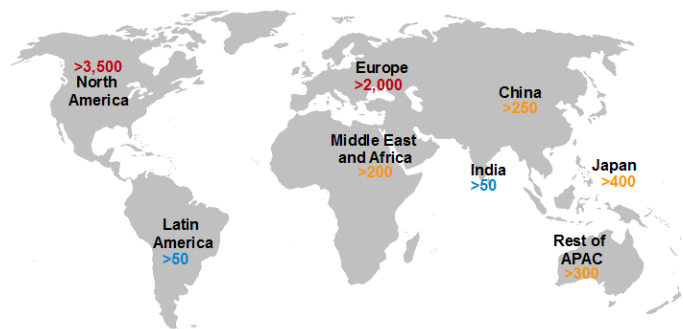

Figure 4: Amount of new data stored varies across geography. New data stored (in Petabytes (PB)) by geography in 2010. New data stored is defined as the amount of available storage used in a given year [9].

Today's supply chain management poses many problems. It performs poorly in responding to market demands, supplier condition, etc. in real-time. The big data collected through IoT infrastructure based on sensing as a service has the potential to diminish these inefficiencies. Research on supply chain management [30] has identified five technologies that will make the process more efficient: mobility, Internet of Things, big data, predictive analysis, and cloud computing. These five areas are strongly interconnected. Internet of Things and mobility provide the sensors that can sense in real-time even while moving. These sensors will produce big data, that is high volume, high variety data in high velocity. So these collected data need to be analysed in order to extract knowledge. Predictive analysis techniques will do the job. However, it cannot be done in traditional computing environment where processing, and storing power are limited and expensive. An elastic infrastructure, such as the cloud, needs to be used to perform such techniques over big data. In other terms, the cloud binds to the Internet of Things [7].

In business perspective, big data has the potential to generate more revenue, reduce risk, and predict future outcomes with greater confidence in low cost [31]. IBM [31] has identified some of the challenges that can be addressed by using big data. For example, analyse multi-channel customer sentiment and experience, detect life-threatening conditions at hospitals in time to intervene, predict weather patterns to plan optimal wind turbine usage, and optimize capital expenditure on asset placement, make risk decisions based on real-time transactional data, identify criminals and threats from disparate video, audio, and data feeds.

The following statistical facts summarise the value of big data [9].
- $300 billion potential annual value to US health care.
- €250 billion potential annual value to Europe's public sector administration.
- $600 billion potential annual consumer surplus from using personal location data globally.
- 60% potential increase in retailers' operating margins possible with big data
- 140,000–190,000 more deep analytical talent positions.
- 1.5 million more data-savvy managers needed to take full advantage of big data in the United States.

4. **Technologies around Big Data**

McKinsey Global Institute [9] discusses some of the existing technologies such as machine learning techniques that need to be extended to be used in big data. Further, McKinsey report identifies additional technologies such as massive parallel-processing (MPP) databases, distributed file systems, cloud computing technologies, etc. that complement the big data management.

The techniques need to be developed to tackle the challenges such as extracting, transformation, integrating, sorting and manipulating data [32]. Basic techniques of meaningful data extraction from big data consists five steps [29]: define, search, transform, entity resolution, answer the query. These conceptual steps are more related to traditional data mining research domain. However, technology behind these conceptual steps will be varying significantly due to the unique characteristics of big data. SAP's Hana tool (sap.com/hana) provides analysis of massive amount of data 3,600 times faster for real-time business.

Scalable and distributed data management has been a key priority among the database research community for more than three decades. Due to lack of cloud feature and high costs in relational database management systems are less favour to be used in big data and cloud environment [33]. NoSQL movement has fuelled the big data significantly. As storage is essential component in big data, there are numbers of commercial and open source solutions available that supports big data requirements. Commercial products are Greenplum (greenplum.com), IBM DB2 (ibm.com/software/data/db2) or Netezza (netezza.com), Microsoft SQL Server (microsoft.com/sqlserver), MySQL (mysql.com), Oracle (oracle.com), or Teradata (teradata.com).

In NoSQL[2] technologies, there are four varieties: Key-value, document store, wide column stores, and graph databases. Out of these four technologies, graph databases are favoured to be used to store information with complex relationships among pieces of data such as social network data, semantic and linked data. Neo4J is an example for this type of database.

---
[2] nosql-database.org

Other three technologies are more favoured to be used commonly across in big data, cloud and IoT applications. Popular key-value based data storage products are Dynamo (aws.amazon.com/dynamodb), Redis (redis.io), Riak (basho.com/Riak), Amazon SimpleDB (aws.amazon.com/simpledb), and Windows Azure Table Storage (windowsazure.com). Popular document store technology based products are CouchDB (couchdb.apache.org), and MongoDB (mongodb.org). Popular wide column based products are Apache HBase (hbase.apache.org), Hadoop (hadoop.apache.org), and Cassandra (cassandra.apache.org). However, the one getting the most attention is Hadoop as it based on popular distributed processing friendly MapReduce[3] technology. However, there is no single perfect data management solution for the cloud to manage big data.

Due to high number of big data producers and high frequency of data generation, the gap between data available to organization and data an organization can process is getting wider all the time as illustrated in figure 5 [5].

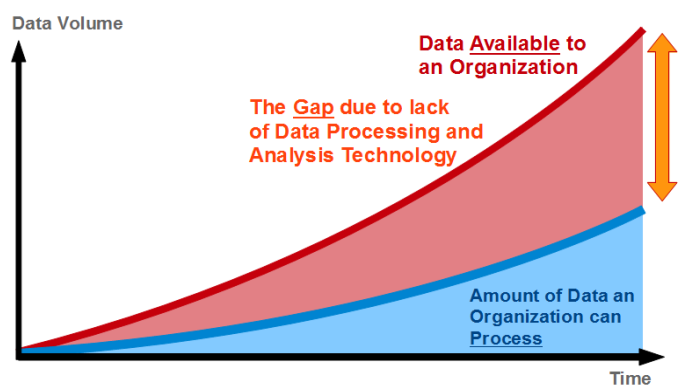

Fig 5: the volumes of data available to organizations today in on rise rapidly, while the volumes of data that can be processed by an organization rise slowly. This creates a gap between the two. In Other words, the percentage (%) of data an organization can analyse is on the decline [5].

Pervasive systems produce high-dimensional data with the use of very sophisticated and advanced sensors. They require high computational power to be processed and storage. For example, high-dimensional data such as brain signal (EEG) are usually processed using client-server architecture. However, [34] has shown that map-reduce based scalable cloud computing architecture can process brain signal more efficiently. In addition, how to manage sensor data in cloud environment has been explained in [35] in relation with wearable sensor domain. Arduino as sensor platform, Google App Engine (appengine.google.com) as cloud platform, and Google Datastore as the database to store sensor data has been employed in the experiments. Connecting and managing sensor via cloud is a critical milestone in the process of sharing sensors and sensor data in Sensing as a service model.

---

[3] http://research.google.com/archive/mapreduce.html

## 5. Sensing as a service Model

Providing everything as a service is model that emerged with cloud computing. Garter defines [10] cloud computing as *a style of computing in which massively scalable IT-related capabilities are provided "as a service" using Internet technologies to multiple external customers.*

Cloud computing is expected to play a significant role in IoT paradigm. The cloud storage and processing capabilities are essential to make the IoT vision a reality.

Cloud computing consists of three main layers or model, Infrastructure as a Service (IaaS), Platform as a Service (PaaS), and Software as a Service (SaaS). Each layer has been discussed comprehensively in [10]. In addition to the above main layer, some other layers are also introduced and discussed in literature such as Database as a Service (DBaaS), Data as a Service (DaaS), Ethernet as a Service (EaaS), Network as a Service (NaaS), Identity and Policy Management as a Service (IPMaaS), and Sensing as a service (SaaS). In general, all these models are called XaaS, which means 'X' can be virtually anything. In this paper we discuss only, Sensing as a service (SaaS) model [36], [37].

IoT envision that sensors to be attached everywhere. In such an environment, owner sill be able publish their sensor data and earn a fee or discount as a return. To visualize in this concept in real world let's consider a scenario.

Mike bought a new refrigerator for his new home. He brought it home and plugged it to the power. Fridge automatically identified the availability of WIFI in the house. Then, prompted Mike for asking permission to be connected to the Internet via the display available in the refrigerator. Once he agreed, refrigerator identified the sensors attached to itself such as RFID reader, temperature, door sensors in refrigerator and deep-freezer, etc. Refrigerator prompted Mike asking whether he would like to publish these sensors on the Internet so interested parties would access them by paying a fee or any other offer. Mike agreed to publish the sensors given that he would return a satisfactory offer. After some time, Mike received an email from a company called *DairyIceCreeam*, an ice cream manufacturer, with and offer. *DairyIceCreeam* is interested to have access to the RFID reader in Mike's refrigerator and the door sensor attached to the deep-freezer. As a return, *DairyIceCreeam* will offer either 5% discount on every product sold by *DairyIceCreeam* or a monthly fee of 2 dollars. As Mike likes *DairyIceCreeam* products, he agreed the offer of 5% discount instead of the monthly fee.

This scenario explains sensing as a service at personal level where similar processes could be applied at public and private organization levels. Owners of sensors (could be a person, private organization, public organization, government, etc) will be able to publish their sensor data and get a return on investment. Privacy and security need to be handled appropriately. Further, at least three categories of people would participate in the process: Sensor owner such as Mike, sensor publishing and managing companies (organizations act as mediators such as eBay), and the sensor data consumers such as *DairyIceCreeam cooperation*. Further, in *DairyIceCreeam* corporate perspective, it is much cheaper for them to acquire sensors as a service so they don't need to deploy sensors themselves or manually collect data from users. It also provides real-time and accurate data which is far better than manually collect data regarding customer behaviour via retail stores or directly via customers. By accessing RFID reader and deep-freezers door sensor, DairyIceCreeam can acquire much more detailed and comprehensive set of information which is impossible to collect using traditional methods. For example, with use of RFID tags in ice cream containers, *DairyIceCreeam* can find out information such as how many times user consumes dairy per week, at which time users are more likely to eat them and so on.

In the above scenario, we considered only one person (Mike) and one thing (refrigerator). As IoT envisioned, when 50 billion things connected to the Internet, they will generate big data at unprecedented scale. Further strengthening our idea IEEE spectrum says *doesn't be surprised when your fridge joins Facebook* [6].

In IoT domain, sensor data are not the only information that will be stored. More importantly relevant context information will be always annotated to the raw sensor data for later retrieval purpose. The context data add more meaning and value to the sensor data. However, it will also increase the storage requirements in many times. For example [38], for each sensor reading context information such as what properties are measured, which sensors available, where are they located, how are they configured, who is responsible will be stored.

Architectural design challenges for sensor web are presented in [39] as flexibility to facilitate different application requirements, distributed reasoning and decision making, number of layers to be average not low or high, ease of deployment, and availability in term of licensing. [39] have also survey variety of different architectures and sensor web applications. Sensor web architecture and application can be considered as preliminary approached of sensing as a service model. SaaS can be developed on top of these sensor web architectures.

Sensor-Cloud [40] is an infrastructure that aims at managing physical sensors by connecting them to the cloud. Sensor-Cloud uses SensorML[4] to describe the metadata of physical sensors such as physical sensors' description and measurement processes. SensorML is a standard model and an XML encoding mechanism for describing sensors defined OGC (Open Geospatial Consortium). Sensor-Cloud bundles the physical sensors in o virtual sensors where users can combine them together to achieve advanced results. Sensor-Cloud allows users to use sensors without worrying about the details. It also provides physical sensor management capabilities such as checking the usage of their physical sensors. However, Sensor-Cloud does not focus on offering sensor data as a service. Instead, it focuses on managing sensors via cloud.

SenaaS [41] (Sensor-as-a-Service) has proposed to encapsulate both physical and virtual sensors in to services according to Service Oriented Architecture (SOA). SenaaS mainly focuses on providing sensor management as a service rather than providing sensor data (collection and dissemination) as a service. It consists three layers: real world access layer, semantic overlay layer, and services virtualization layer. Real world access layer is responsible for communicating with sensor hardware. Semantic overlay layer adds semantic annotation to the sensor configuration process. Services virtualization layer facilitates the users. Similar approach is presented in [42] where sensors are wrapped in to service in SOA with semantic annotation which leads to automated sensor service discovery and composition

OpenIoT [43] (Open Source blueprint for large scale self-organizing cloud environments for IoT applications) is aimed at developing an open source middleware platform to connect Internet-objects to the cloud. The OpenIoT project will leverage existing IoT middleware solutions such as Global Sensor Network [5] and ASPIRE (fp7-aspire.eu). OpenIoT defines IoT as a combination of sensor data and applications. OpenIoT concentrates on providing a cloud-based middleware infrastructure in order to deliver on-demand access to IoT services, which could be formulated over multiple infrastructure providers [44]. The Objective of IoT is to provide utility based services such as Sensing as a service, Location-as-a-Service, and Traceability-as-a-Service.

Every year, Australian grain breeders plant up to 1 million 10 m2 plots across the country to find the best high yielding varieties of wheat and barley. The plots are usually located in remote places – often requiring more than four hours travel one-way to reach. The challenge is to monitor the crop performance and growing environment throughout the season and return the information in an easily accessible format. "Phenonet" wireless sensor networks are now routinely deployed in wheat variety trials throughout Australia by staff at the High Resolution Plant Phenomics Centre (HRPPC) (http://phenonet.com) in order to collect information over a field of experimental crops. Currently, the sensors are deployed around 40 plots. These sensors produce two million data points per week. If and when the sensors are eventually deployed around all or majority of plots, the overall amount of sensed data will really be big data.

---

[4] www.opengeospatial.org/standards/sensorml

[5] sourceforge.net/apps/trac/gsn/

## 6. Challenges in Big Data Management

The challenges in big data can be broadly divided in to two categories: engineering and semantic [29]. Engineering challenge is to perform data management activities such as query, and storage efficiently. Semantic challenge is to extract the meaning of the information from massive volumes of unstructured dirty data. Several other challenges in big data are presented in [29] with details.

The Jet Propulsion Laboratory (JPL) has identified number of major challenges in big data management [17].

- **High volume of processing using low power consumed digital processing architecture.** Power that needs to process the data as well as power that is required to cool the processing system need to be considered together in designing such systems.
- **Discovery of data-adaptive Machine learning techniques that can analyse data in real-time.** These techniques are critical specifically, in situation where data, high time and spectral resolution, are produced by many antennas in Very Long Baseline Array that cannot be stored due to size for later analysis.
- **Design scalable data storages that provide efficient data mining.** Storing any kind of data is useless unless they cannot be retrieved and extract the knowledge efficiently. For example, if we send all the data sensed to the cloud to be stored at very high sampling rate, the wide-area networks may cause network congestion. They are always a trade of between when we decide where to store and process, specifically in IoT domain. If we choose to process in the cloud, all the data need to be send to the cloud which will resulted unacceptable communication latency. Similarly, if we choose to process data locally, the resource may not be enough to achieve the desired result at desired speed.

Patidar et al [10] have identified a number of key challenges in big data management related to the cloud.

- **Data security and privacy**: This is about managing access control to the big data. It is critical in Sensing as a service model that we discussed earlier.
- **Approximate results**: Due to the volume and the velocity of big data, approximate results would be order of magnitude faster compared to traditional query execution. It need be decided where and when to use approximation where it will not harm the accuracy of the results or decision that emerged based on approximated results.
- **Data exploration to enable deep analytics**: Certainly, new technologies are required to analyse large volumes of data faster with efficient resource and power consumptions.
- **Enterprise data enrichment with web and social media**: The big data has more relationships among themselves that ever in history. This is also related to Linked Data and social media. Real value of big data will be merged once we are able to preserver and identify the relationships between data.
- **Query optimization**: In order to harvest the knowledge hidden in big data optimised query processing in essential. Optimization need to consider in different perspectives such as energy consumption, required memory, processing time, storage requirements, etc. Parallel processing would be the key in query processing in cloud environments.
- **Performance isolation for multi-tenancy**: This is about designing and developing model of performance service level agreements for multi-tenant data systems that can be metered at low overhead.

## 7. Conclusions

Big data consists of hidden gold (high-valued data) mixed with dirty (noise, erroneous and raw) data. The advanced systems and innovative technologies will allow us to efficiently process massive amounts of dirty data and extract gold out of it. It will require concentrated efforts by ICT researchers and practitioners to meet the challenge of big data and ride the wave of information implosion.

## Acknowledgment

Support from EU FP7 OpenIoT project and SSN TCP, CSIRO is thankfully acknowledged.